\documentclass[aps,prl,twocolumn,showpacs]{revtex4}
\usepackage{graphicx} 
\def\gapx{\lower 2pt \hbox{$\buildrel>\over{\scriptstyle{\sim}}$\ }}
\def\lapx{\lower 2pt \hbox{$\buildrel<\over{\scriptstyle{\sim}}$\ }}
\def\ph2{{\it p}-H$_2$}
\def\beq{\begin{equation}}
\def\eeq{\end{equation}}
\def\Am3{\AA$^{-3}$}
\begin{document}
\title{Thin helium film on a glass substrate}
\author{Massimo Boninsegni         }
\affiliation{Department of Physics, University of Alberta, Edmonton, Canada T6G 2G7}

\date{\today}

\begin{abstract}
We investigate by Monte Carlo simulations the structure, energetics and superfluid properties of  thin $^4$He films (up to four layers) on a glass substrate, at low temperature. The first adsorbed layer is found to be solid and ``inert", i.e., atoms are localized and do not participate to quantum exchanges. Additional layers are liquid, with no clear layer separation above the second one. It is found that a single $^3$He impurity resides on the outmost layer, not significantly further away from the substrate than $^4$He atoms on the same layer.
\end{abstract}
\maketitle

\section{Introduction}
\label{intro}
Adsorption of thin films of the most abundant isotope of He ($^4$He) on various substrates, is a subject of longstanding interest, motivated by the unusual, often intriguing physical properties that these films display \cite{bruch97}. A chief example is the rich phase diagram of Helium on graphite, which has been the subject of an intense investigation over the past three decades. Significant attention has also been devoted to adsorption  on weakly attractive substrates, e.g., those of alkali metals, following original suggestions  that, at low temperature, the weakness of the physisorption potential ought to result in unusual wetting \cite{cheng91,cheng93} and superfluid \cite{boninsegni99} behavior.

Less is known quantitatively, both theoretically and experimentally, about adsorption over a glass substrate, despite its obvious relevance to experimental studies of the superfluid properties of liquid (and solid) helium embedded in such porous materials such as vycor or aerogel \cite{reppy92}. Of particular interest are issues such as the interplay between the one or more solid, ``inert" layers that are expected to form over the substrate, and the liquid-like behaviour that should progressively emerge as successive layers are adsorbed. Glass is the archetypal disordered substrate, and that renders its theoretical modelling particularly challenging.

Microscopic theoretical studies of adsorption of helium (both pure \cite{apaja03b,buffoni05} and in isotopic mixtures \cite{pricaupenko95}) on glass, based on realistic interaction potentials, have typically made use of a vary simple model of the substrate, described as a flat surface on account of its lack of a definite crystal structure. As mentioned above, the substrate is in reality irregular, disordered, and this can be expected to promote the appearance of a stable, low-coverage insulating film, in which atoms are localized at randomly located wells of the adsorption potential; such a phase is obviously lost if a flat substrate model is assumed. As the coverage is increased, however, a first-order transition to a regular (monolayer) crystal is expected, and as further layers are adsorbed the corrugation of the substrate ought to become less relevant, i.e., a flat substrate model should capture the essential physics of the problem (this is what observed, e.g., in numerical simulations of helium films on regular crystalline substrates, such as graphite \cite{corboz08}). 

In this work we study the adsorption of $^4$He on a glass substrate, modelled as in all previous studies by others, by means of numerical simulations based on the continuous-space Worm Algorithm (WA). The purpose of this investigation is that of determining quantitatively structural and superfluid properties of the adsorbed $^4$He film at low temperature. Specifically, we present results for $T$=1 K and $T$=0.5 K. We considered films up to four layers thick, and also looked at a single $^3$He impurity diluted in the film, for different values of coverage. There is considerable uncertainty as to the actual values of the parameters that are used in the potential describing the interaction of a helium atom with the substrate, and we attempted to pick values that overlap with most previous studies.

Our main findings are the following:
\begin{enumerate}

\item{The first helium adlayer is solid, whereas the successive ones are liquid. In the solid ``inert"  layer, atoms are highly localized, the layer is fairly well separated from the second, and quantum exchanges (both intra- and inter-layer) are virtually absent. On the other hand, successive layers are liquid-like, with no sharp separation between layers above the second. Atomic exchanges take place, mostly within the layer for the second layer, as well as inter-layer above the second one. Layers above the first are all superfluid in the low temperature limit. Thus, one can conceptually distinguish the first adlayer from the rest of the system. Previous studies \cite{pricaupenko95} have often assumed that two inert layers would form, but the results presented here show that the model potential utilized is not sufficiently strong to stabilize more than one such layer.}

\item{A single $^3$He impurity diluted in the film, for different values of $^4$He coverage, always positions itself on the top layer. Its wave function is not significantly more spread out than that of a $^4$He atom on the same layer, i.e., the $^3$He atom does not spend a significantly fraction of time at greater distances from the substrate. In other words, $^3$He atoms do not ``float" on top of the $^4$He film, as one might have expected.}

\end{enumerate}

The remainder of this manuscript is organized as follows: in the next section, the model Hamiltonian is briefly introduced, and some details of the computational method utilized  are presented (for a thorough  illustration of the methodology, the reader is referred to Refs. \onlinecite{worm,worm2}); we then discuss our results and outline our conclusions in the two following sections.

\section{Model}
\label{model}
Our system of interest is modelled as an ensemble of $N$ $^4$He atoms, regarded as point particles,  moving in the presence of an infinite, smooth planar substrate (positioned at $z$=0). The system is enclosed in a vessel shaped as a parallelepiped, with periodic boundary conditions in all directions \cite{note}. Let $A$ be the area of the substrate; correspondingly, the nominal $^4$He coverage is $\theta=N/A$. The quantum-mechanical many-body  Hamiltonian  is the following:
\begin{equation}\label{one}
\hat H = -{\lambda}\sum_{i=1}^N \nabla_i^2 + \sum_{i<j} V(r_{ij}) +\sum_{i=1}^N U(z_i)\ .
\end{equation}
 Here, $\lambda$=6.0596 K\AA$^2$, $V$ is the potential describing the interaction between two helium atoms,  only depending on their relative distance, whereas $U$ is the potential describing the interaction of a helium atom with the substrate,  also depending only on the distance of the atom from the substrate. We use the accepted Aziz potential \cite{aziz79} to describe the interaction of two Helium atoms, which has been used in almost all previous studies, and has also been shown \cite{worm2} to provide an accurate description of the energetics and structural properties of liquid $^4$He in the superfluid phase.  

More delicate is the issue of which potential to use to describe the interaction of a helium atom with the smooth substrate (i.e., the $U$ term in (\ref{one})). The simplest potential is the so-called ``3-9":
\begin{equation}\label{39}
U_{3-9}(z) = \frac{D}{2}\biggl ( \frac {a^9}{z^9}-3\frac{a^3}{z^3}\biggr )
\end{equation}
which is a functional form obtained by integrating the Lennard-Jones potential over a semi-infinite, continuous slab. The parameters $a$ and $D$ are normally adjusted to fit the results of some {\it ab initio} electronic structure calculations for the specific adatom-substrate system of interest. $D$ is the characteristic depth of the attractive well of the potential, whereas $a$ is essentially the location of the minimum of such a well.

\begin{figure}
 \includegraphics[height=3.4in, angle=-90]{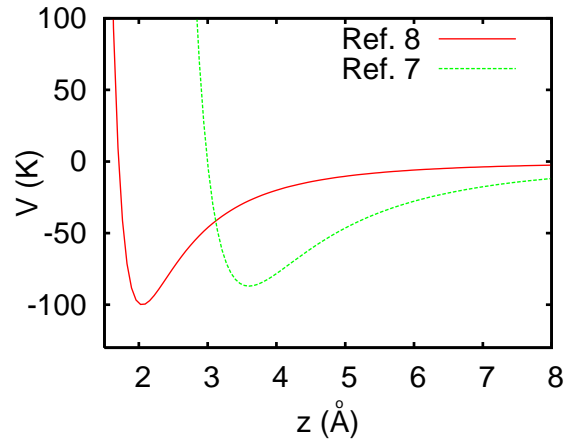}
\caption{Comparison of the helium-glass adsorption potential as given by expression \ref{39} with the parameters provided in Ref. \onlinecite{pricaupenko95} (solid line) and \onlinecite{buffoni05} (dashed line).}
\label{fig:1}       
\end{figure}

There appears to be some ambiguity as to the values of the coefficients that would make this potential appropriate for the system of interest here. 
Choices made in some of the previous studies are summarized in Table \ref{tab:1}. Although these studies were aimed at describing the environment experienced by a helium atom in the confines of various porous materials, such as vycor, the ``building block" for all of them is the interaction potential felt by a helium atom close to an infinite, flat glass (SiO$_2$) substrate. As is seen in the table, there seems to be a reasonable agreement over the value of $D$, which is $\sim$ 100 K in all studies. On the other hand, while  $a$ has been taken to be slightly above 2 \AA\   in Refs. \onlinecite{apaja03b,pricaupenko95},  in Ref. \onlinecite{buffoni05}  it was set to a much larger value, namely 3.6 \AA.  As shown in Fig. \ref{fig:1}, the ensuing potentials are quantitatively very different, the parameters utilized in Ref. \onlinecite{buffoni05} giving rise to a potential with a significantly more attractive long-range tail (in turn likely to give rise to quantitatively different adsorption characteristics). 
\\
The claim is made in Ref. \onlinecite{buffoni05} that $a$=3.6 \AA\ applies to the interaction of helium atoms with a substrate of MgO (regarded as sufficiently similar to SiO$_2$); but the most reliable estimates reported in the literature  (see, e.g., Ref. \onlinecite{vidali}) place $a$ for such a system in the 2.2-2.5 \AA\ range. In this work, we utilize the same values of $a$ and $D$ taken in Ref. \onlinecite{pricaupenko95}, which are close to those of Ref. \onlinecite{apaja03b}.
\begin{table}
\caption{Values of the coefficients $a$ and $D$ that appear in the ``3-9" potential describing the interaction between a helium atom and a glass substrate, utilized in previous studies.}
\label{tab:1}       
\begin{tabular}{lll}
\hline\noalign{\smallskip}
Ref. & $a$ (\AA) & $D$ (K)  \\
\noalign{\smallskip}\hline\noalign{\smallskip}
\onlinecite{apaja03b} & 2.18 & 128 \\
\onlinecite{buffoni05} & 3.6 & 87 \\
\onlinecite{pricaupenko95} & 2.05 & 100 \\
\noalign{\smallskip}\hline
\end{tabular}
\end{table}

\section{Methodology}
We performed Quantum Monte Carlo (QMC) simulations of the system described 
by  Eq. (\ref{one}) using the continuous-space Worm Algorithm. This methodology has
emerged over the past few years as a far superior alternative to conventional Path Integral
Monte Carlo (PIMC), to compute accurate thermodynamic properties of Bose systems at finite temperature. Specifically, the WA allows one to simulate systems comprising up to two orders of magnitude more particles than PIMC, allowing for a reliable extrapolation of values of physical observables to the thermodynamic limit.
\\
The specific implementation 
utilized in this project is {\it canonical}, i.e., we keep the number $N$ 
of particles fixed \cite{fabio}.  Other technical aspects of the calculations 
are common to any other  QMC simulation scheme. 

\section{Results}
\subsection{Computational details}
Results were obtained by 
simulating systems comprising a number $N$ of particles ranging from 36 to 108, 
spanning a range of $^4$He coverage $\theta$ from 0.050 to 0.300 \AA$^{-2}$. Estimates 
presented here are obtained at two temperatures, namely $T$=1 K and $T$=0.5 K; based 
on the wealth of accumulated theoretical and experimental data for liquid $^4$He, 
$T$=1 K is 
low enough that structural and energetic properties are essentially those of the ground state.
The usual fourth-order high-temperature propagator utilized in all previous
studies based on the WA was adopted here; convergence of 
the estimates was observed for a time step 
$\tau\approx 3.125\times 10^{-3}$ K$^{-1}$, which corresponds to 
$P$=320 imaginary time ``slices" at  $T$=1 K. 
Because the computational cost was negligible, all estimates reported 
here were obtained using twice as small a time step, in order to be on 
the safe side. 
\begin{figure}
\includegraphics[height=3.4in, angle=-90]{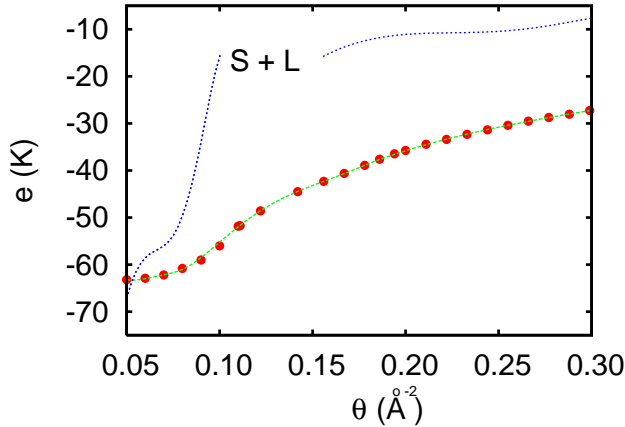}
\caption{Energy per particle in the low-temperature limit for an adsorbed $^4$He film on a glass substrate, computed by QMC simulations (circles),  as a function of coverage. Statistical errors are smaller than the symbol size. Dashed line through circles is a polynomial fit to the data. Dotted line is the chemical potential, obtained from the fit. The label S+L indicates a region of liquid puddles on top of the first solid layer.}
\label{fig:2}       
\end{figure}
\subsection{Energetics}
Fig. \ref{fig:2} shows the energy per particle $e(\theta)$ as a function of coverage, computed by QMC simulations at $T$=1 K. Assuming these to be essentially ground state results, based on a piecewise polynomial fit to the estimates for $e(\theta)$ and the relation
\begin{equation}
\mu(\theta) = e(\theta) + \theta \frac {de}{d\theta}
\end{equation}
we obtained the $T$=0 chemical potential, also shown in Fig. \ref{fig:2}.
\\
The monolayer equilibrium coverage  $\theta_e$, for which $\mu(\theta_e) = e(\theta_e)$, is found to be  close to 0.052 \AA$^{-2}$, to be compared to 0.043 \AA$^{-2}$ for the purely two-dimensional system \cite{giorgini}. This corresponds to a superfluid monolayer, but it is important to restate that the model of substrate considered here lacks the corrugation and disorder of the surface, which will stabilize a disordered, ``glassy" insulating monolayer in the real system.
On increasing coverage, the system forms a two-dimensional (triangular) crystal at $\theta_{1C}\approx$ 0.08 \AA$^{-2}$. The crystal can be compressed up to $\theta_{1P}$=0.100 \AA$^{-2}$, at which second layer promotion is observed.  The system forms a stable liquid second layer at $\theta_{2L}$=0.153  \AA$^{-2}$, whereas in the coverage region $\theta_{1P} \le \theta\le \theta_{2L}$ puddles of liquid form on top of the solid first layer.
\\
Above $\theta_{2L}$, the chemical potential is a smooth, monotonically increasing function of coverage, approaching asymptotically the value $\sim$ -7.2 K for bulk liquid $^4$He. This signals continuous growth of the adsorbed film, without well-defined layer formation. This is confirmed by direct observation of the structure of the film, as well as by its superfluid properties.
\subsection{Structure}
\begin{figure}
\includegraphics[height=3.4in, angle=-90]{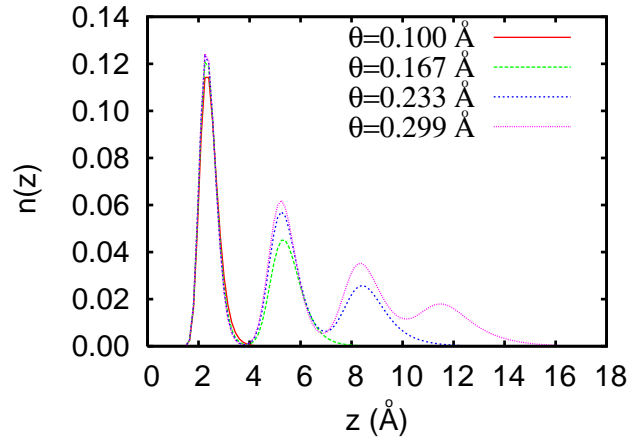}
\caption{Density profiles $n(z)$ in \AA$^{-3}$, computed by QMC, in the direction perpendicular to the substrate, at various coverages. The temperature is $T$=1 K.}
\label{fig:3}       
\end{figure}
%
%
Fig. \ref{fig:3} shows the computed $^4$He density profiles $n(z)$, in the direction perpendicular to the substrate, for four different coverages. The first observation is that there exists a fairly sharp demarcation between the innermost (first) layer and the second, with essentially no overlap between the two, and no interlayer atomic exchange. On the other hand, second and third layers already overlap significantly, and essentially no separation exists between third and fourth layers. The fourth layer is broad and ill-defined, its average density very close to that of equilibrium liquid $^4$He (0.022 \AA$^{-3}$).
\begin{figure}
\includegraphics[height=3.2in, angle=-90]{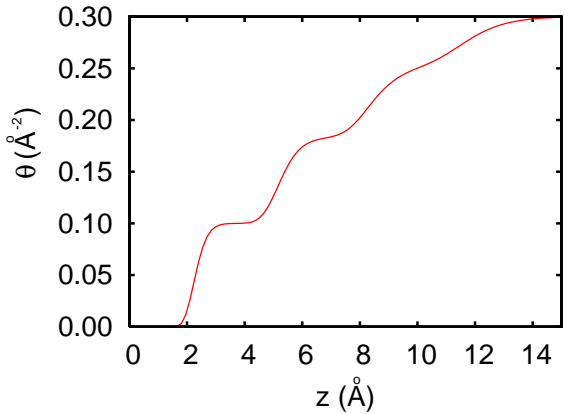}
\caption{Integrated coverage $\theta(z)$ (in \AA$^{-2}$), defined in Eq. \ref{inte}, as a function of the distance $z$ from the substrate, for a film of total coverage 0.299 \AA$^{-2}$. The temperature is $T$=1 K.  }
\label{integ}       
\end{figure}

Additional evidence of the ill-defined character of layers in this system is provided in Fig. \ref{integ}, where the integrated coverage 
\beq\label{inte}
\theta(z)\equiv \int dx^\prime\ dy^\prime\ \int_0^z dz^\prime \rho(x^\prime,y^\prime,z^\prime)
\eeq
which depends on the distance from the substrate,  is shown for a film of total coverage $\theta$=0.299 \AA$^{-2}$.
Steps can be seen in correspondence of the appearance of the second layer (at approximately 0.100 \AA$^{-2}$) and of the third at roughly 0.165 \AA$^{-2}$ can be clearly seen. However, the boundary between third and fourth layer is already scarcely visible. This is not qualitatively dissimilar from what is observed on other substrates \cite{pavloff}. or The wealth of theoretical data accumulated over a wide variety of substrates, generally suggests that although the details of the first layer clearly reflect the properties of the underlying substrate (strength and corrugation), beyond the first layer what one observes seems to be essentially substrate-independent.

Atoms in the first layer are arranged on a regular triangular crystal and are not involved in permutational exchanges, either within the layer itself or with atoms of the second layer. Both the first and second layers are slightly compressed as further layers form on top. At its highest compression, the second layer has a half-width at half maximum around 1 \AA, and corresponds to an effective coverage of approximately 0.050 \AA$^{-2}$, i.e., the second layer is already liquid. This contradicts the assumption of two solid adsorbed layers of helium on the substrate, modelled as in this work (with the same choice of parameters), made in Ref. \onlinecite{pricaupenko95}, as the potential (\ref{39}) with the choice of parameters made here, only leads to the formation of a single solid layer.
\subsection{Superfluidity}
Unlike structure and energetics, the superfluid properties of a systems studied by computer simulation are strongly affected by its finite size, i.e., the number of particles. In particular, phase transitions are smeared, and one will typically observe a finite value of the superfluid fraction $\rho_S$ for temperatures above the critical temperature $T_c$.  

In this study, given the crudeness of the model utilized we are not interested in providing accurate values of the transition temperature, but rather to assess reliably whether an adsorbed film of a given coverage can be expected to display superfluidity below a reasonably well-defined temperature.
Because of the two-dimensional (2D) nature of the system studied here, the superfluid transition is expected to fall in the Kosterlitz-Thouless universality class \cite{kt78}, with a sharp universal ``step" at $T_c$.  Finite-size effects affect mostly estimates obtained in the vicinity of $T_c$, whereas for $T < T_c$, the results for a finite system are quantitatively representative of the thermodynamic limit, where $\rho_S \sim \rho_S(T=0)$ [unity for a translationally invariant system].
\begin{figure}
 \includegraphics[height=3.4in, angle=-90]{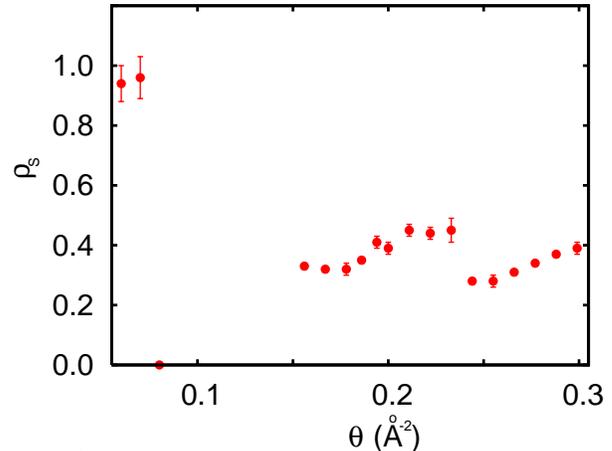}
\caption{$^4$He superfluid fraction $\rho_S$ of adsorbed films of various coverages $\theta$ (in \AA$^{-2}$), computed by QMC simulations. The temperature is $T$=0.5 K.}
\label{fig:4}       
\end{figure}

The most recent estimate for $T_c^{2D}$ is 0.655 K, based on the Aziz pair potential utilized here, which affords fairly closed agreement with experiment for bulk liquid $^4$He \cite{worm,worm2}.
Computer simulations of an adsorbed $^4$He monolayer on Lithium \cite{boninsegni99}, yielded evidence that such a film essentially mimics the physics of 2D $^4$He, and because the substrate considered here is considerably more attractive than Lithium, we can expect the same to be true here.
Indeed, the superfluid fraction of the equilibrium stable monolayer film at $T$=0.5 K, is close to 100\%, consistently with its behaviour being fairly close to that of purely two-dimensional $^4$He (such a monolayer is considerably narrower than on a Li substrate, see Ref. \onlinecite{boninsegni99}).   

\begin{figure}
\includegraphics[height=3.4in, angle=-90]{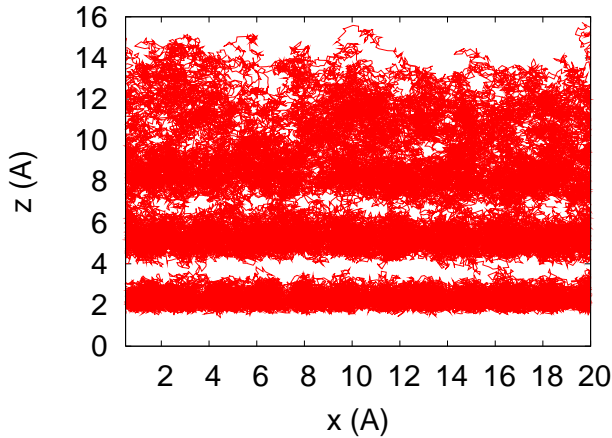}
\caption{Snapshot of an instantaneous many-atom configuration for a $^4$He film of coverage $\theta$=0.299 \AA$^{-2}$ at $T$=0.5 K. Shown are particle world lines projected on the $xz$ plane. Clearly identifiable are only the two bottom layers, the second being liquid. Quantum many-particle exchanges involving atoms in the second layer are largely restricted to that layer.}
\label{fig:snap}
\end{figure}

As mentioned above, as the coverage is increased the first layer crystallizes, and superfluidity disappears, to reappear only when a stable liquid second layer forms, at a coverage $\theta_{2L}$. As shown in  Fig. \ref{fig:4}, the superfluid fraction at $\theta=\theta_{2L}$ at $T$=0.5 K is close to 33\%, i.e., consistent with an underlying non-superfluid crystalline layer, and an entirely superfluid second layer on top. That particles in the first layer are not involved in quantum-mechanical exchanges is inferred by histograms of permutation cycles, as well as from the lack of overlap between the first and second layer, as shown in Figs. \ref{fig:3} and \ref{fig:snap}.
\\
Given that no evidence is observed of crystallization for layers above the second, and that the superfluid transition temperature is expected to be {\it generally} an increasing function of the film thickness (the superfluid transition temperature should approach the bulk value $T$=2.177 K in the thick film limit), it seems reasonable to posit that $T$=0.5 K should be sufficiently low to render finite-size effects essentially unnoticeable.

At coverages above $\theta_{2L}$, quantum exchanges involving particles in the same layer (other than the first)  occur at $T$=1 K already, interlayer exchanges being more frequent between the third and fourth layer, which overlap most substantially (as shown in Fig. \ref{fig:snap}). Consequently, the superfluid fraction is finite for any stable film consisting of more than one layer, confirming that all layers above the first are liquid. Results for the $^4$He superfluid fraction $\rho_S$ at $T$=0.5 K are shown in Fig. \ref{fig:4}. The behaviour as a function of $\theta$, for the coverages above $\theta_{2L}$ is oscillatory, with local minima indicating layer ``completion". However, $\rho_S$ stays finite at all coverages. As the coverage increases, the superfluid fraction is expected to approach 100\% in the thick film limit.

\subsection{Single $^3$He impurity.}
\begin{figure}
\includegraphics[width=0.40\textwidth]{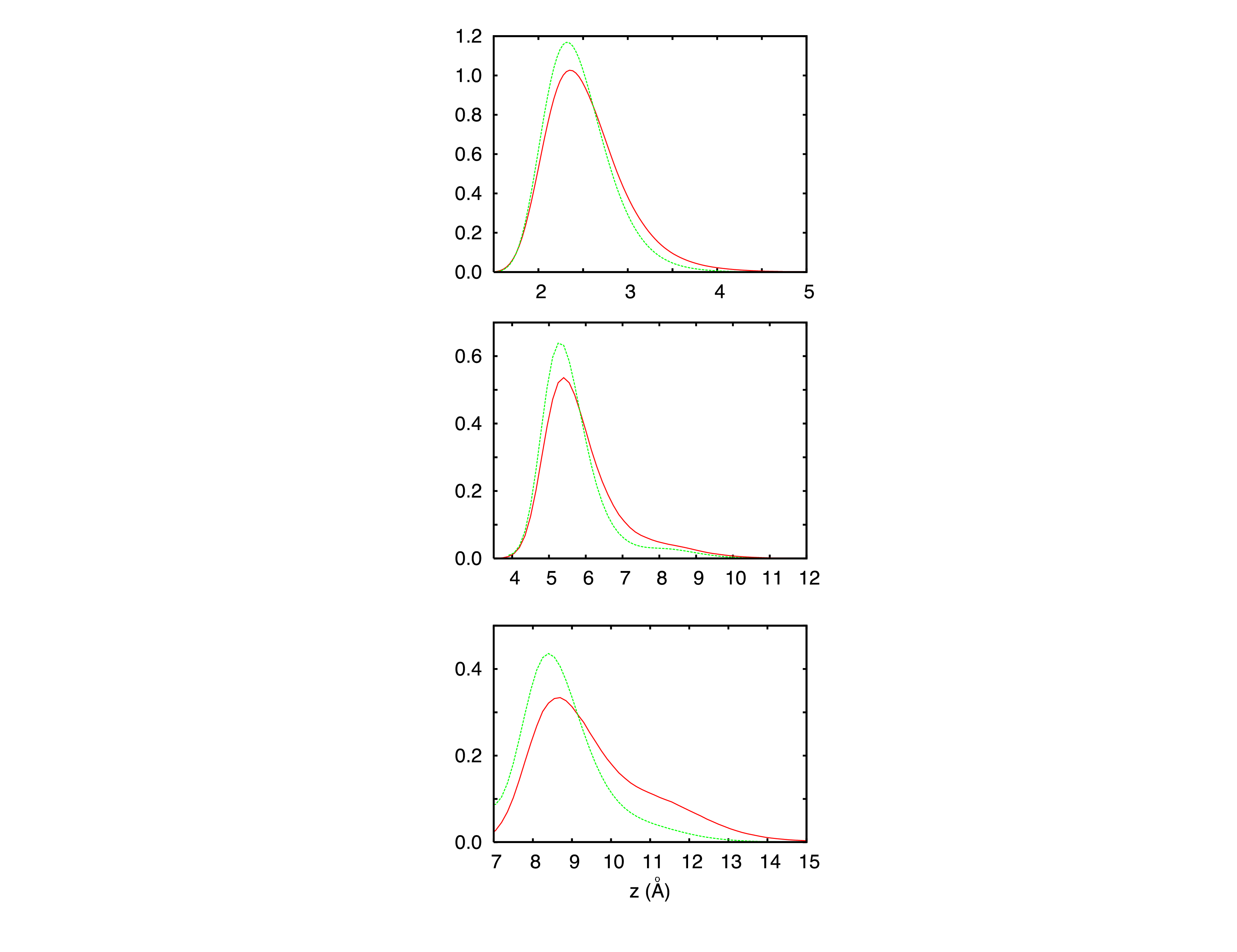}
\caption{Probability density of position in the direction perpendicular to the substrate, for a $^3$He atom embedded in a  $^4$He film of coverage $\theta$=0.100 \AA$^{-2}$ (top, single layer), 0.178 \AA$^{-2}$ (middle, two layers) and 0.244 \AA$^{-2}$ (bottom, three layers). Dashed lines show the corresponding probability density for a $^4$He atom in the same layer where the $^3$He atom resides, which is always the top one. Units on the ordinate axis are arbitrary.}
\label{fig:5}
\end{figure}

The physics of a single $^3$He atom dissolved in an adsorbed $^4$He film is of interest for a number of reasons. First of all, $^4$He is never isotopically pure, as contaminating $^3$He impurities are always present and their effect on the behaviour of $^4$He in confinement, or restricted geometries, is not completely understood. Moreover, isotopic liquid helium mixtures are of considerable fundamental interest, as one of the simplest Fermi-Bose mixtures accessible in nature. Their phase diagram has been extensively investigated experimentally, including in the confines of porous glasses such as aerogel \cite{kim93}.

The most basic question that one can ask, is in what region of an adsorbed $^4$He film of varying thickness will a $^3$He atom position itself. The wave function of a single $^3$He atom embedded in an adsorbed  $^4$He monolayer is expected to be more ``spread out", i.e., the atom is expected spend more time further away from the substrate than a $^4$He, because of its lighter mass. In the case of a few adsorbed $^4$He layers, it is presumed that the $^3$He atom will reside in the top layer, near the free surface of the film, where the three-dimensional density approaches that of bulk liquid $^4$He. Conceivably, the $^3$He atom might even ``float" on top of the outmost $^4$He layer.

In order to assess these issues quantitatively, we have carried out simulations of a single $^4$He impurity at the same helium coverages considered for the case of a pure $^4$He film. At all coverages explored here, the $^3$He atom is found to reside on the top layer of the film. Fig. \ref{fig:5} shows the probability density of position in the direction perpendicular to the substrate, for a single $^3$He atom embedded in adsorbed $^4$He films of various coverages, corresponding to one, two and three adsorbed layers. 
\\
In all cases shown, the $^3$He and $^4$He density profiles overlap substantially, to indicate that the $^3$He impurity is essentially embedded in the $^4$He layer; it does not, e.g., float atop. On average,  the lighter atom lays only very slightly further away from the substrate than a $^4$He atom on the same layer, although this displacement becomes more pronounced as the number of adlayers increases. This is attributable to  the weakening of the attraction exerted by the substrate on the $^3$He atom (laying near the surface) as the film thickens, and the enhanced quantum delocalization of $^3$He with respect to atoms of the heavier isotope. In the limit of a thick $^4$He film, the $^3$He impurity is expected to remain near the surface, where the presence of a bound state has been predicted theoretically \cite{saam,cohen73,epstein} and verified experimentally \cite{guo, gasparini}.

\section{Conclusions}
We have studied by accurate, first principles Quantum Monte Carlo simulations the physics of an adsorbed film of $^4$He over a model glass substrate, at low temperature. We utilized the most commonly adopted model of a glass substrate, neglecting the disorder and irregularity of its surface. 
 We considered films up to four layers thick, and also looked at a single $^3$He impurity diluted in the film, for different values of coverage. 
\\
It is found that only the first adlayer is solid, whereas the successive ones are liquid. The solid ``inert"  layer  layer is fairly well separated from the second, atoms are highly localized and no quantum exchanges (both intra- and inter-layer) take place. On the other hand, successive layers are liquid-like, with no sharp separation between layers above the third. In the second layers, quantum exchanges are mainly confined to the layer itself, whereas interlayer exchanges occur between third and fourth layers.
\\
A single $^3$He impurity is always observed on the top layer, not near the substrate. This   appears to be in contrast with the conclusion of Ref. \cite{pricaupenko95}, where the existence of a homogeneous $^3$He-$^4$He mixture was predicted near the substrate, based on a density functional calculation. On the other hand, the results shown here lend credence to the scenario proposed in Ref. \cite{boninsegni95}, according to which the heavier isotope in the mixture, $^4$He, would be prevalently found in the vicinity of the surface of, e.g., aerogel strands.   The wave function of the $^3$He atom, however, is not significantly more spread out than that of a $^4$He atom on the same layer, i.e., the $^3$He atom does not spend a significantly fraction of time at greater distances from the substrate. 

\section*{Acknowledgements}
This work was supported in part by the Natural Science and Engineering Research Council of Canada under research grant 121210893, and by the Alberta Informatics Circle of Research Excellence (iCore).


\begin{thebibliography}{}
\bibitem{note}
{The size of the box in the direction perpendicular to the substrate, is chosen sufficiently large to render the presence of periodic boundary conditions in this direction immaterial.}
\end{thebibliography}


\end{document}